\documentclass[a4paper,11pt]{article}
\pdfoutput=1 

\usepackage{jcappub} 

\usepackage[T1]{fontenc} 

\title{QSOs sigposting cluster size halos as gravitational lenses: halo mass, projected mass density profile and concentration at $z\sim0.7$}

\author[a]{L. Bonavera\note{Corresponding author.},}
\author[a]{J. Gonz\'{a}lez-Nuevo,}
\author[b]{S.L. Su\'{a}rez G\'{o}mez,}
\author[c,d,e]{A. Lapi,}
\author[f]{F. Bianchini,}
\author[g]{M. Negrello,}
\author[h]{E. D\'{i}ez Alonso,}
\author[a]{J. D. Santos,}
\author[h]{F. J. de Cos Juez}

\affiliation[a]{Departamento de F\'{i}sica, Universidad de Oviedo, C. Federico Garc\'{i}a Lorca 18, 33007 Oviedo, Spain}
\affiliation[b]{Departamento de Matem\'{a}ticas, Universidad de Oviedo, C. Federico Garc\'{i}a Lorca 18, 33007 Oviedo, Spain}
\affiliation[c]{SISSA, Via Bonomea 265, I-34136 Trieste, Italy}
\affiliation[d]{INFN-Sezione di Trieste, via Valerio 2, 34127 Trieste, Italy}
\affiliation[e]{INAF-Osservatorio Astronomico di Trieste, via Tiepolo 11, 34131 Trieste, Italy}
\affiliation[f]{School of Physics, University of Melbourne, Parkville, VIC 3010, Australia}
\affiliation[g]{School of Physics and Astronomy, Cardiff University, The Parade, Cardiff CF24 3AA, U.K.}
\affiliation[h]{Departamento de Explotaci\'{o}n y Prospecci\'{o}n de Minas, Universidad de Oviedo, Oviedo, 33004 Asturias, Spain}

\emailAdd{bonaveralaura@uniovi.es}
\emailAdd{gnuevo@uniovi.es}
\emailAdd{suarezsergio@uniovi.es}
\emailAdd{lapi@sissa.it}
\emailAdd{federico.bianchini@unimelb.edu.au}
\emailAdd{NegrelloM@cardiff.ac.uk}
\emailAdd{diezenrique@uniovi.es}
\emailAdd{jdsantos@uniovi.es}
\emailAdd{fjcos@uniovi.es}

\abstract{Magnification bias is a gravitational lensing effect that is normally overlooked because it is considered sub-optimal in comparison with the lensing shear. Thanks to the demonstrated optimal characteristics of the sub-millimetre galaxies (SMGs) for lensing analysis, in this work we were able to measure the magnification bias produced by a sample of QSOs acting as lenses, $0.2<z<1.0$, on the SMGs observed by Herschel at $1.2<z<4.0$. Two different methodologies were successfully applied: the traditional cross-correlation function approach and the Davis-Peebles estimator through stacking technique. The second one was found to be more robust for analysing the strong lensing regime ($<20-30$ arcsec in our case) and provides the possibility to take into account the positional errors of the sources in our samples.

From the halo modelling of the cross-correlation function, the halo mass where the QSOs acting as lenses are located was estimated to be greater than $\log_{10}{(M_{min}/M_\odot)} > 13.6_{-0.4}^{+0.9}$, also confirmed by the mass density profile analysis ($M_{200c}\sim 10^{14} M_\odot$). These mass values indicate that we are observing the lensing effect of a cluster size halo signposted by the QSOs, as in previous studies of the magnification bias.

Moreover, we were able to estimate the lensing convergence, $\kappa(\theta)$, for our magnification bias measurements down to a few kpcs. The derived mass density profile is in good agreement with a Navarro-Frank-White (NFW) profile. We also attempt an estimation of the halo mass and the concentration parameters, obtaining $M_{NFW}=1.0^{+0.4}_{-0.2}\times10^{14} M_\odot$ and $C=3.5_{-0.3}^{+0.5}$. This concentration value is rather low and it would indicate that the cluster halos around these QSOs are unrelaxed. However, higher concentration values still provides a compatible fit to the data.}

\begin{document}
\maketitle
\flushbottom

\section{Introduction}
\label{sec:intro}

The gravitational lensing effect is produced whenever a foreground object (lens) magnifies the light rays coming from background sources (\textit{magnification}, $\mu$) and stretches the area of the surrounding sky region (\textit{dilution}). Both effects shift the source number counts of the lensed objects, depending mainly on the counts' slope ($\beta$; $N(>S)= N_0 S^{-\beta}$). As a consequence, the gravitational lensing increases the detection probability of amplified background sources when dealing with a flux-limited sample (``magnification bias'' see, e.g., \cite{Sch92}). It produces an excess/deficit of background sources nearby the lens position. 

On one hand, the strong gravitational lensing happens when the magnification factor is high, $\mu \gtrsim 2$ (implying high matter over-densities), which is easier to detect but very rare. On the other hand, the weak lensing effect, characterised by $\mu \lesssim 2$  and caused by the more common low densities cosmic structures, is more likely to happen and it is the source of most of the magnification bias. As for the magnification bias, the sensitivity is highly enhanced when the background sources are characterised by very steep source number counts, $\beta >2$. Moreover, the magnification bias implies a non zero signal when cross-correlating two source samples with non-overlapping redshift distributions: the lensing effect makes different objects appear spatially correlated (see \cite{Scr05,Men10,Hil13,Bar01}, and references therein).

A new population of galaxies was discovered within Herschel Astrophysical Terahertz Large Area Survey (H-ATLAS; \cite{Eal10}) data, called Sub-Millimetre Galaxies (hereafter SMGs) whose properties (steep source number counts, $\beta > 3$, high redshift, $z>1$, and very low cross-contamination with respect to the optical band) make them the optimal sample for magnification bias studies (see e.g. \cite{GN14, GN17}). In fact, in \cite{GN14} they were able to measure (with high significance, $>10\sigma$) the angular cross-correlation function (hereafter CCF) between selected H-ATLAS high-z sources, z > 1.5, and two optical samples with redshifts 0.2 < z < 0.8, extracted from the Sloan Digital Sky Survey (SDSS; \cite{Ahn12}) and Galaxy and Mass Assembly (GAMA, \cite{Dri11}) surveys. \cite{GN17} (hereafter GN17) constituted a substantial improvement over the CCF measurements made by the previous work, with updated catalogues and wider area (with S/N > 5 above 10 arcmin and reaching S/N$\sim20$ below 30 arcsec). Thanks to the better statistics it was possible to split the sample in different redshift bins and to perform a tomographic analysis (with S/N > 3 above 10 arcmin and reaching S/N$\sim 15$ below 30 arcsec). Moreover, a Halo model was implemented to extract astrophysical information about the background galaxies and the deflectors that are producing the lensing link between the foreground (lenses) and background (sources) samples. In the case of the sources, it was found typical mass values in agreement with previous studies. However, the lenses are massive galaxies or even galaxy groups/clusters, with minimum mass of $M\gtrsim 10^{13} M_\odot$.

On the other hand, quasi-stellar objects or QSOs are extremely luminous active galactic nucleus (AGN). Thanks to their high luminosity, they can be detected over a very broad range of distances making them the perfect background sources for gravitational lensing events. In fact, The SDSS Quasar Lens Search (SQLS) identified 28 galaxy-scale multiply-imaged quasars \cite{Ogu06,Ogu08}. They are QSOs whose light undergoes gravitational lensing, resulting in double, triple or quadruple images of the same QSO. Moreover, QSOs have been used in several cross-correlation studies, but usually as background sources (see for example, \cite{Bar93,Scr05,Men10}). The aim of this work, instead, is to study the QSOs acting as lenses on the H-ATLAS background sample and extract information on the mass density profile and compare it with current theoretical ones.

It should be noticed that identify QSOs acting as lenses is not an easy task and there is very few literature on the subject. It is worth mentioning that \cite{COU10} presents the first detection of strong gravitational lensing caused by a QSO (SDSS J0013+1523 at z = 0.120), by looking for emission lines redshifted behind QSOs in the SDSS spectra (7th data release). In this case the total radial mass profile is not constrained, claiming that for a detailed analysis deep optical HST imaging are needed. 
Moreover, \cite{COU12} reported three new cases of the same kind (SDSS J0827+5224 at z = 0.293, SDSS J0919+2720 at z = 0.209, SDSS J1005+4016 at z = 0.230), whose lensing nature was confirmed thanks to Keck imaging and spectroscopy and HST imaging. 
Later \cite{Har15} measured (by means of weak leansing) the total mass and mass profile of non-obscured, low redshift quasars showing recent merger activity. 
More recently, in the ALMA-LESS survey identified SMGs, \cite{Dan17} found that the ALESS006.1 appears to be lensed by an adjacent low-redshift QSO. Furthermore, a tentative selection of 12 QSOs acting as lenses using the spectroscopic technique has been made by \cite{Mey17}, although not yet published. The candidates are selected in the SDSS-III Data Release 12 within the Baryon Oscillation Spectroscopic Survey (BOSS), and if confirmed may quadruple the number of known QSOs acting as strong lenses.
Probably in the future the number of known QSOs lenses will increase, but getting a statistically significant number of quasar will be very difficult. For this reason we decided to undertake the cross-correlation approach in order to get information on the mass density profile for such objects.

In addition, due to the faint character but high probability of the weak lensing we took advantage of the stacking technique and applied it to this purpose. The aim is to enhance the effects of weak lensing at the expanses of studying the single event (more unlikely to be detected). In fact, the stacking consists in co-adding the signal from many weak or undetected objects to get an overall statistical detection. Among many other applications, such technique has been already conveniently used to fully exploit the Planck data to recover the very weak ISW signal by looking at the positions of positive and/or negative peaks in the gravitational potential (\cite{Pla14,Pla16b}). To study the faint polarised signal of radio and infrared sources detected in total flux density by NVSS and by Planck (see \cite{Stil14}, \cite{Bon17a} for radio and \cite{Bon17b} for infrared sources). To obtain the mean spectral energy distribution (SED) of a sample of optically selected quasars \cite{Bia19}. To detect weak gravitational lensing of the cosmic microwave background at the location of the WISE$\times$SuperCOSMOS galaxies using the publicly available Planck lensing convergence map \cite{Bia18}. Or to probe star formation in the dense environments of $z\sim 1$ lensing haloes aligned with dusty star-forming galaxies detected with the South Pole Telescope \cite{Wel16}.

In this work, we apply the stacking technique and compare its results with the traditional CCF estimator and we use them to study the average total halo mass and the average radial mass density profile associate to dark matter halos of the QSOs acting as lenses. The work is organised as follows. Section \ref{sec:data} describes the data, while section \ref{sec:method} gives details of the methodology applied and the results obtained with the stacking approach and the traditional cross-correlation estimator. In section \ref{sec:denprof} our results are used to estimate the mass density profile of the QSOs as lenses and in section \ref{sec:concl} our conclusions are drawn. 

\begin{figure}
\centering
\includegraphics[width=\linewidth]{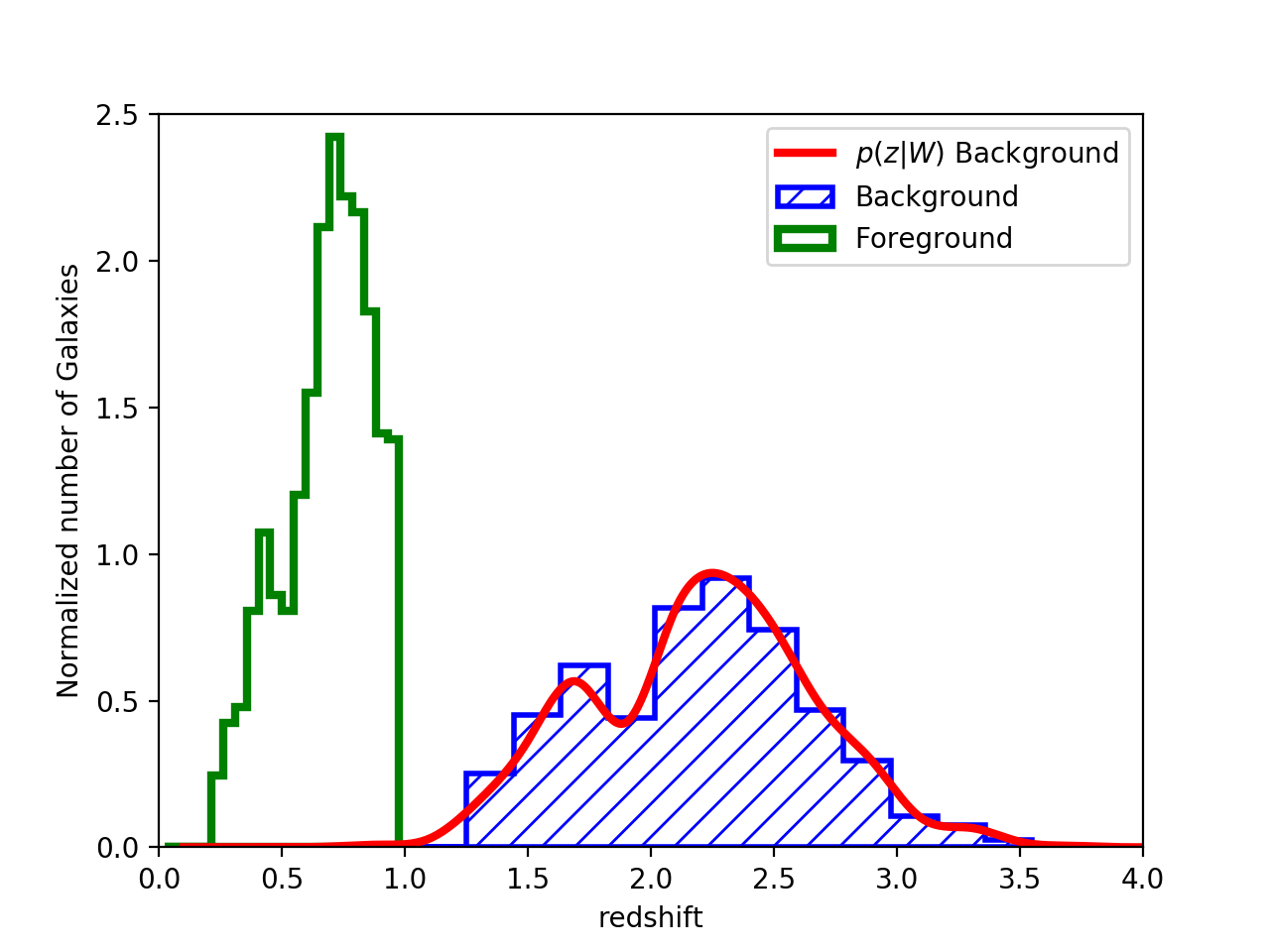}
\caption{\label{fig:zhists} Redshift distributions of the background H-ATLAS sample (blue histogram) and the foreground QSOs one (green histogram). The estimated p(z|W) of the background sample taking into account the window function and the photometric redshift errors is represented as a red line.}
\end{figure}

Throughout the paper, a flat $\Lambda CDM$ cosmology has been adopted with the best-fit cosmological parameters determined by \cite{Pla16a,Pla18}: matter density $\Omega_m$ = 0.31, $\sigma_8$ = 0.81 and Hubble constant $h = H_0 /100$ $km$ $s^{-1} Mpc^{-1} = 0.67$.

\section{Data}
\label{sec:data} 

The background sample has been selected from the H-ATLAS data, the largest area extragalactic survey carried out by the Herschel space observatory \cite{Pil10}. With its two instruments PACS \cite{Pog10} and SPIRE \cite{Gri10} operating between 100 and 500 $\mu$m, it covers about 610 deg$^{2}$. 
The survey is comprised of five different fields, three of which are located on the celestial equator (GAMA fields or Data Delivery 1 (DR1); \cite{Val16,Bou16,Rig11,Pas11,Iba10}) covering in total an area of 161.6 deg$^2$. The other two fields are centred on the North and South Galactic Poles (NGP and SGP fields or DR2; \cite{Smi17,Mad18}) covering areas of 180.1 deg$^2$ and 317.6 deg$^2$, respectively. We used as background sources the officially detected galaxies in the three H-ATLAS GAMA fields that correspond to equatorial regions at 9, 12 and 14.5 h, and only the NGP region ones from the DR2.

In both H-ATLAS DRs there is an implicit $4\sigma$ detection limit at 250 $\mu$m ($\sim S_{250} > 29$ mJy). The estimated $1\sigma$ total noise level (both instrumental and confusion) for source detection at 250 $\mu m$ is 7.4 mJy for both DR1 and DR2 catalogues \cite{Val16,Mad18}. In addition, following GN17, a $3\sigma$ limit at 350 $\mu$m has been applied to increase the robustness of the photometric redshift estimation.

Due to the scanning strategy there are small overlapped regions, surveyed more than 2 times, where the noise properties are different. In \cite{Amv19}, the potential effect of these non-uniformities on the measurement of the auto-correlation function was studied in detail and it was demonstrated negligible. The main reason is because the total noise level is shown to be mainly dominated by the confusion noise produced by the faint non-detected galaxies that is uniform along the fields. In any case, this kind of non-uniform noise distribution has no relevant effect when studying pairs of galaxies from two different catalogues within a certain angular distance: it will increase the number of pairs, at maximum.

Moreover, only sources with photometric redshift between 1.2 and 4.0 have been taken into account to ensure that there is no overlap in the redshift distribution of lenses and background sources. The photometric redshifts were estimated by means of a minimum $\chi^2$ fit of a template SED to the SPIRE data (using PACS data when possible). It was shown that a good template is the SED of SMM J2135-0102 (`The Cosmic Eyelash' at z = 2.3; \cite{Ivi10,Swi10}), that was found to be the best overall template with $\Delta z/(1 + z) = -0.07$ and a dispersion of 0.153 \citep{Ivi16,GN12,Lap11}.
In the end, we are left with 57930 sources that constitute approximately the 24 per cent of the initial sources.

The redshift distribution of the background sample is shown in Fig. \ref{fig:zhists} (blue hashed histogram). The mean redshift of the sample is $\left< z\right> = 2.2_{-0.5}^{+0.4}$ (the uncertainty indicates the $1\sigma$ limits).
To allow for the effect on the redshift distribution of random errors in photometric redshifts, as in GN17, we estimate the redshift distribution, $p(z|W)$ (red line in Fig. \ref{fig:zhists}), of galaxies selected by our window function, a top-hat for $1.2 < z < 4.0$.

Our QSO initial sample is obtained from the one used in \cite{Bia19}. It was selected from the publicly available SDSS-II and SDSS-III Baryon Oscillation Spectroscopic Survey (BOSS) catalogues of spectroscopically confirmed QSOs detected over 9376 $\deg^2$.
In particular, we make use of the QSO catalogs from the seventh (DR7, \cite{Sch10})\footnote{Available at \url{http://classic.sdss.org/dr7/products/value_added/qsocat_dr7.html}.} and twelfth SDSS data releases (DR12, \cite{Par17})\footnote{Available at \url{http://www.sdss.org/dr12/algorithms/boss-dr12-quasar-catalog/}.}. 
We refer the reader to \cite{Ros12} for a discussion on the details of the QSO target selection process.
Here we simply recall that whereas the DR7 catalog mostly includes ``low-$z$'' sources at $z < 2.5$, the DR12 sample specifically targeted QSOs at $z > 2.15$ \cite{Par17}). 
However, as a consequence of a color degeneracy in the target selection from photometric data, a fraction of lower redshift QSOs has been re-observed, resulting in a secondary maximum of the QSO redshift distribution around $z \simeq 0.8$.
A merged sample was created by combining the QSO catalogues from the DR7 and DR12 catalogs. 
Any duplicate is removed by retaining the QSO information contained in the DR12 sample.
As we are interested in studying the QSOs acting as the lenses on the background sample, and in order to minimised the potential cross-contamination, we selected only QSOs with redshift between $0.2<z<1.0$ (1546 in the common area). Their redshift distribution is shown in Fig. \ref{fig:zhists} (green histogram) with $\left< z\right> = 0.7_{-0.2}^{+0.1}$ (the uncertainty indicates the $1\sigma$ limits).

In this work, even taking into account the background photometric uncertainties, we considered the possible cross-contamination (sources at lower redshift, z < 1.0, with photometric redshifts > 1.2) statistically negligible. This conclusion was addressed in more detail in previous works (see \cite{Lap11, GN12, GN14, GN17}) with different kind of independent tests. Moreover, in \cite{Bia19} it was also verified that only $\sim2\%$ of the QSOs are detected by Herschel. This fact reduces even more the potential cross-contamination between foreground and background samples.

\begin{figure*}
\centering
\includegraphics[width=.45\textwidth]{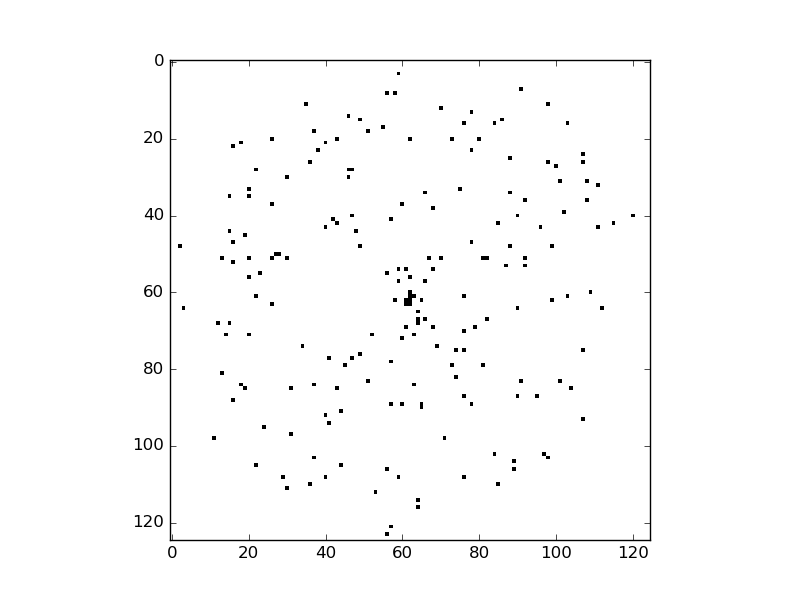}
\hfill
\includegraphics[width=.45\textwidth]{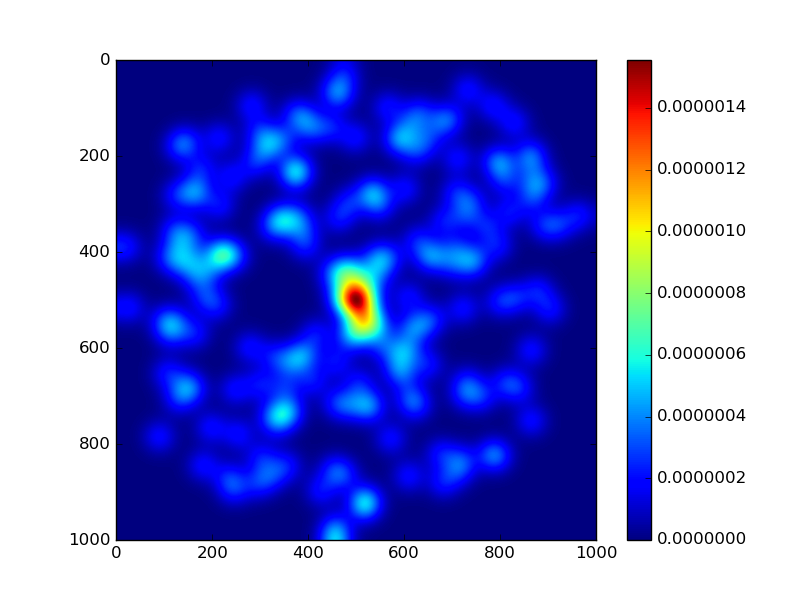}\\

\includegraphics[width=.45\textwidth]{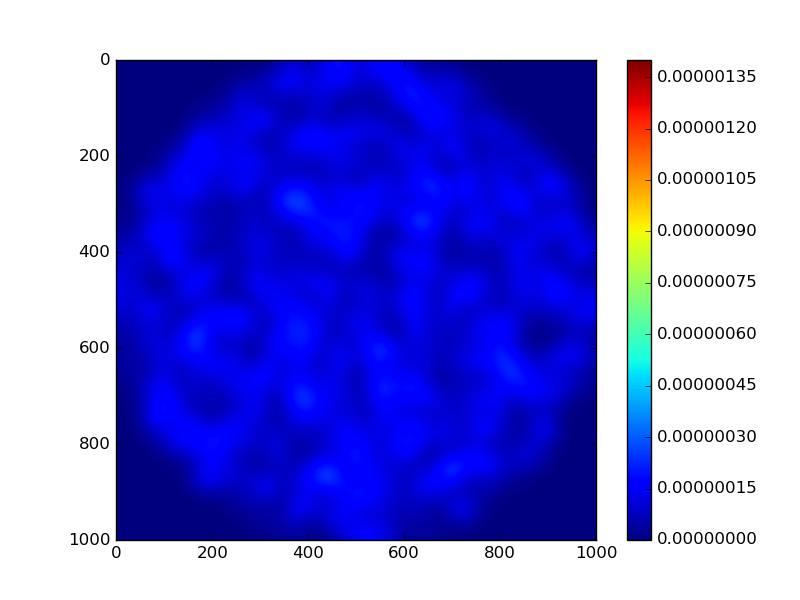}
\caption{Top Left: Stacked image of the QG pairs, 209, in grey scale ($125\times125$ pixels) using a pixel size of 0.8 arcsec and without filtering. Top Right: stacked image ($1000\times1000$ pixels) using a pixel size of 0.1 arcsec and applying a $2.4\sigma$ beam gaussian filter (see text for more details). Bottom: stacked image of the random data for the same radius and pixel size as for the actual data and filtered with the same gaussian filter.}
 \label{fig:analysis_4Z}
\end{figure*}  

\section{Measurements}
\label{sec:method} 

\subsection{Stacking}
\label{sec:stack} 

Stacking is a statistical method that consists in adding up many regions of the sky centred in previously selected positions (see \cite{Dol06, Mar09, Bet12} and references therein). In this way, the noise/background signal can be reduced, since it is expected to randomly fluctuate about the mean value: by adding up random higher and lower values with respect to the mean value, the signal (the mean itself) is enhanced. This technique is useful when the signal to be measured is too weak (e.g. sources in the sample are too faint or \textit{the number of the events per lens are too low}, as for the current work), but a reasonably large number of events is available. So, stacking can provide overall statistical information for the target signal when a high S/N ratio cannot be achieved with a single event.

In our case, the magnification bias produces an excess of sources detected just above the detection limit thanks to the (\textit{small}) gravitational lensing amplification near the galaxies acting as lenses. Therefore, we are interested in the excess of detected sources within a certain angular separation with respect to the random scenario. For this reason, we are stacking the positions of the background sources not their flux density, i.e. we only use the H-ATLAS catalogues of detected galaxies and not the maps. 
This is the same starting point as for the traditional CCF estimator, however, the stacking approach has some advantages, such as the possibility to take into account the positional errors for the sources (not possible in the traditional CCF estimator) and to follow the pairs contributing to the final stacked map. This pair information become very useful to estimate, with better precision, important physical quantities as the critical surface density (see Sec. \ref{sec:theory} and \ref{sec:constrains}).

\begin{figure}
\centering
\includegraphics[width=\linewidth]{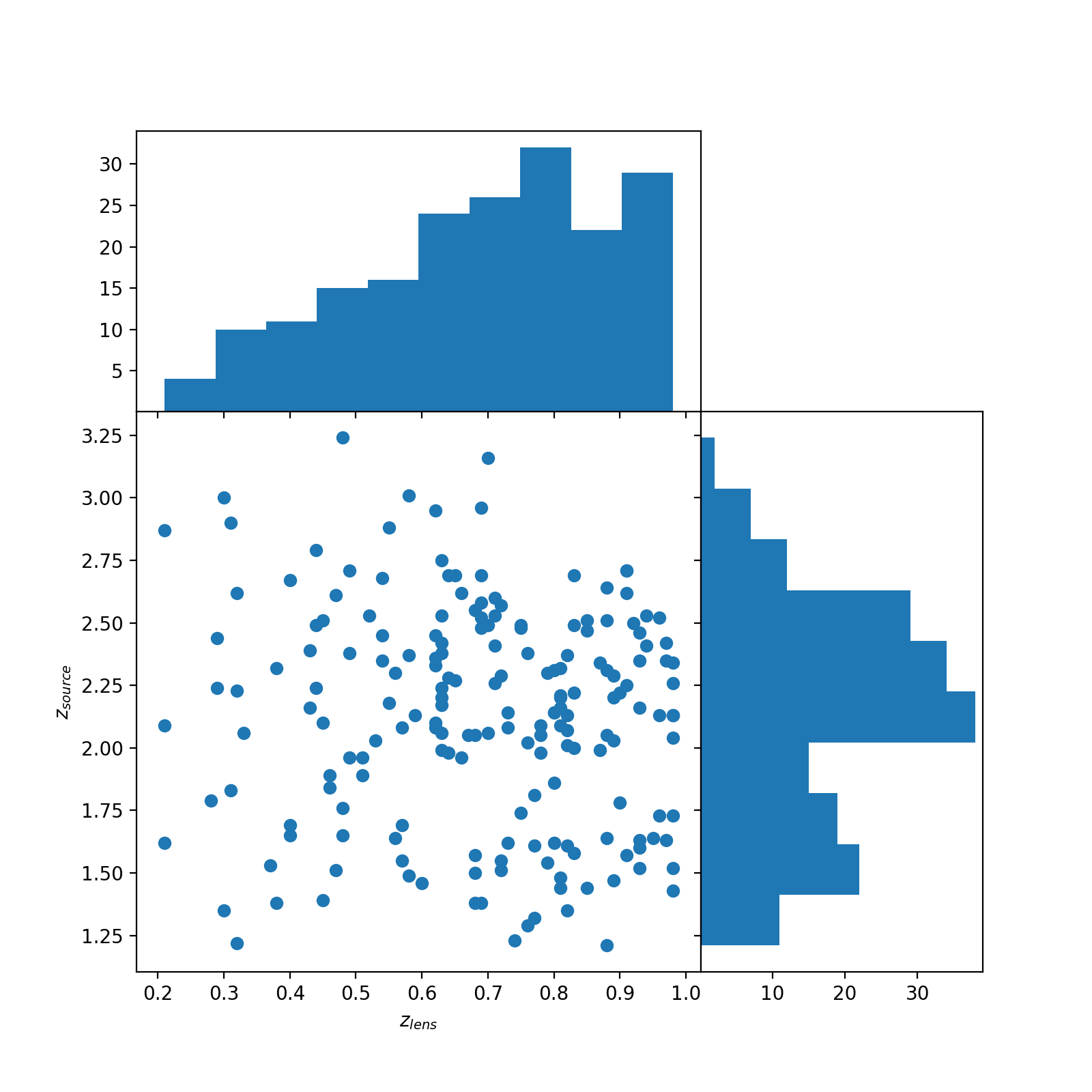}
 \caption{Redshifts of the QG pairs found with an angular separation lower than 50 arcsec. These are the QG pairs used to estimate the stacked image shown in Fig. \ref{fig:analysis_4Z}.}
 \label{fig:pairs_dndz}
\end{figure}

To derive the stacked magnification bias produced by the QSOs acting as lenses over the background SMGs, we searched for the sources in the background sample that fall within the region centred in the lens position and within a distance given by $r = n_{pix}\times pixsize/2$, where $n_{pix}$ is the length in pixels of the patch and $pixsize$ is the size of the pixels (in arcsec). In this way we obtain a map of $n_{pix} \times n_{pix}$ pixels centred in the lens position containing the nearest background sources of the lenses (QG pairs). 

We repeated this procedure for all the sources in the lenses sample and add all the maps to obtain the stacked map (normalised to the number of the maps that have been added, 1546 targets in total). In order to take into account the positional errors in the catalogues, we applied a Gaussian filter of $\sigma$=2.4 arcsec to the map, the positional accuracy estimated for the H-ATLAS catalogues \cite{Bou16,Mad18}. This step is equivalent to change every background position of a QG pair for a 2D isotropic gaussian describing the probability that the background source is not exactly in the catalogued position but in a near location (the positional uncertainty).
The positional uncertainty for the lenses is of the order of the pixel size and, therefore, considered negligible for our analysis.

In principle, such stacked map can be built with any pixel size. As both the foreground and background samples are built from different catalogues observed in different wavelengths bands with almost negligible cross-contamination emission between both bands, there is no intrinsic resolution limit. This is a clear advantage with respect to these studies that use both samples observed in the optical band, were the lens galaxy make impossible to observed the phenomena at angular scales smaller than its size.  For our purposes, we choose a patch size of $n_{pix}$=1000 and $pixsize=$0.1 arcsec that allow us to study both the weak and strong lensing regimes.

The resulting map with the identified QG pairs is plotted in the top panels of Fig. \ref{fig:analysis_4Z}. The total number of QG pairs is 209: 45 in G09, 34 in G12, 44 in G15 and 86 in NGP. They show an almost isotropic, but not completely homogeneous, distribution of QG pairs at an angular distance lower than 50 arcsec. Although most of the signal is produced by the weak lensing effect, close to the centre, we can locate a region with much higher density, i.e. higher lensing probability. As discussed in more detail later, this stronger excess of QG pairs below 10-20 arcsec is due to the strong lensing effect. This first panel shows just the position of the QG pairs and the number and size of the pixels (125 pixels of 0.8 arcsec each) has been chosen in order to make such pairs visible to human eye on the plot, same reason as for the grey-scale. The excess of QG pairs is more clearly shown in the top right panel, that correspond to the same map but with the actual number and size of pixels used for the analysis (1000 pixels with a size of 0.1 arcsec) and smoothed to take into account the positional errors as described earlier. As shown in Fig. \ref{fig:pairs_dndz}, the redshift distributions of both the lenses and sources of the identified QG pairs are very similar with respect the one from their parent samples (see Fig. \ref{fig:zhists}). This means that there is no evident selection bias within our QG pair sample.

The bottom panel of Fig. \ref{fig:analysis_4Z} is used to compare the observed map with what would be the signal in absence of lensing. Simulated random QG pairs distribution maps are produced by creating random lenses source catalogues and applying the same pipeline as for the data. Although negligible, in this way the random realisations take into account the same non-uniform noise distribution of the background sample (as discussed in Sec. \ref{sec:data}).
In order to gather homogeneity in the random map, we simulated 3000 targets for each GAMA region and 7000 for the NGP region (ten times the round-up value of the available foreground possible lenses). 
This random map is needed for the posterior analysis of the measured signal and it is useful to demonstrate that our QG pairs distribution map shows the actual signal due to lensing effects. 

\begin{figure*}
\centering
\includegraphics[width=\linewidth]{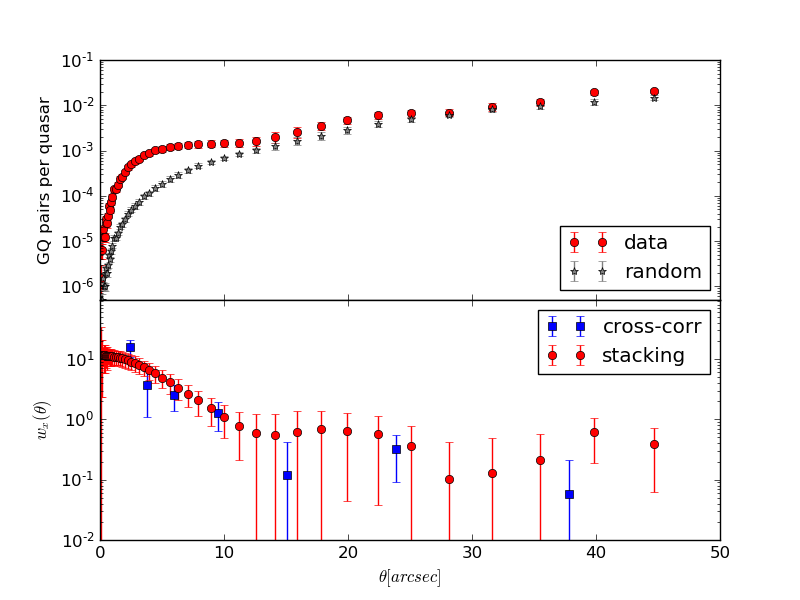}
 \caption{Analysis of the stacked data on all the available regions, with a pixel size of 0.1 arcsec. The top panel shows the results for the circular analysis of the stacked data (red circles) and of the simulated random data (grey stars); the bottom panel shows the results once applied eq. \ref{eq:xcor_cfr} (red circles) and by directly estimating the CCF as in GN17 (blue squares).}
 \label{fig:results_npix100}
\end{figure*} 

\subsection{Cross-correlation function estimation}
\label{sec:xcorr}
Not only is the QG pairs distribution map interesting by itself, but it is also useful to estimate the CCF of the analysed catalogues in order to compare it with previous published results and extract physical information from the lensing system or, in our case, the lens population.

Therefore, we estimated the CCF by drawing concentric circles centred at the patch centre whose radius increases with logarithmic steps of 0.005 arcsec (starting from 0.1 arcsec).
We added all the values of the pixels included in each circular crown defined by these circles, except for the first one that define a circle (not a crown) of 1 arcsec radius. We applied this procedure to the QG pairs distribution map obtained with both real and simulated random data. Then, we computed the quantity given by eq. \ref{eq:xcor_cfr} (the \textit{Davis-Peebles standard estimator}, \cite{DAV83}) for all the bins:
\begin{equation}
\label{eq:xcor_cfr}
	w_x(\theta)=\frac{DD}{RR}-1
\end{equation}
where DD stands for the data and RR for the random realisations. This expression for $w_x(\theta)$ is the analogue to the angular cross-correlation, i.e. the excess of probability to find a QG pair with respect to a random distribution (see below).

To compute the errors both for real data and the random simulations, we perform a Jackknife analysis in each circular crown, by dividing it in ten sub-sectors. We use the Jackknife standard errors as uncertainties for our estimated values. These errors are then propagated to obtain the error-bars of the values obtained using eq. \ref{eq:xcor_cfr}. Moreover, we also consider the potential correlation introduced by the positional uncertainties by computing the covariance matrix when performing the Jackknife analysis. Such covariance matrix has then been taken into account when performing the fit to compare the data with the models.

On the other hand, we also applied the traditional methodology to estimate the CCF, as in \cite{GN14} or \cite{GN17}. We computed the cross-correlation between our background and foreground samples using a modified version of the Landy-Szalay estimator \cite{HER01,Lan93}:
\begin{equation}
\label{eq:xcor}
w_x(\theta)=\frac{\rm{D}_1\rm{D}_2-\rm{D}_1\rm{R}_2-\rm{D}_2\rm{R}_1+\rm{R}_1\rm{R}_2}{\rm{R}_1\rm{R}_2}
\end{equation}
where $\rm{D}_1\rm{D}_2$, $\rm{D}_1\rm{R}_2$, $\rm{D}_2\rm{R}_1$ and $\rm{R}_1\rm{R}_2$ are the normalised data1-data2, data1-random2, data2-random1 and random1-random2 pair counts for a given separation $\theta$. Data1 and data2 refer to the background sources and lens catalogues, and random1 and random2 to their random simulations. 
We have adopted the same procedure as in GN17 by computing the angular CCF in mini-regions and estimating the mean values and their associated standard errors. The ``integral constraint'' correction is considered negligible due to the relative large area of each mini-region ($\sim 14\, \rm{deg}^2$).

Fig. \ref{fig:results_npix100} shows our results for both methodologies. The top panel shows the estimated radial profile for actual data (red circles) and random data (grey stars). The bottom panel shows the quantity obtained by applying eq. \ref{eq:xcor_cfr} to our results (red circles) and the comparison with those coming from the direct estimation counting QG pairs (blue squares). At scales above $\gtrsim15$ arcsec the lack of enough QG pairs produces an inhomogeneous distribution that is translated into bigger error bars and fluctuating behaviour. At small scales, the effect of the smoothing is clearly seen in the stacking approach. If the positional uncertainty were not taken into account, the measured CCF would diverge due to the steep decline of the random probability at small angular scales. If a QG pair is measured closer than the real distance, the estimated CCF will be much higher than the real one. If the QG pair is instead measured farther than the real distance, the effect is less important becoming negligible above a few arcsec. This bias starts to be noticeable in the first blue square at around $\sim 3$ arcsec and it is unavoidable unless the positional uncertainty is taken into account in the CCF estimator (not possible with the traditional approaches).
Overall, this two sets of results are in good agreement confirming that they are describing the same physical quantity.

From this comparison we can conclude that the stacking technique provides more robust information at the small scales (< 20-30 arcsec) with respect to the traditional CCF. Moreover, the stacking approach is also able to take into account the positional uncertainties of the instruments that are crucial to obtain precise information at such small scales. For this reason, it should be the preferred methodology when performing such analysis in the so called strong lensing regime. However, at larger scales the necessity of better target statistics and the arising of additional issues, as those due to survey borders, make the traditional CCF estimation a better choice.

\begin{figure}
\centering
\includegraphics[width=\linewidth]{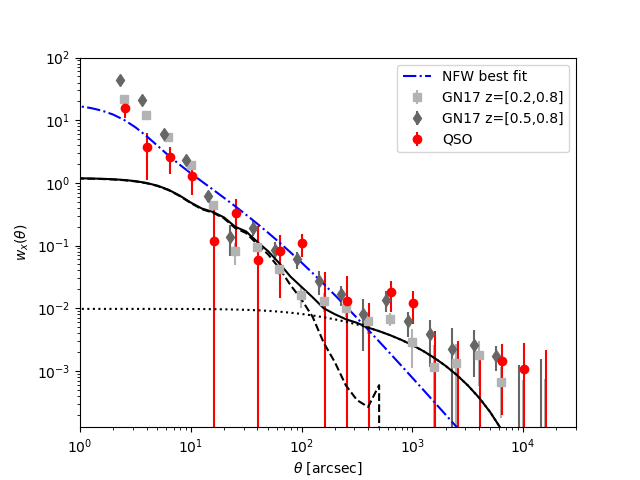}
 \caption{CCF computed following GN17 on the QSOs sample presented in this work (red circles), compared with the one from previous results by GN17 obtained using elliptical galaxies in the $0.2-0.8$ (grey squares) and $0.5-0.8$ (dark grey diamonds) redshift ranges. The best fit halo model is shown by the black solid line, being the black dotted and dashed lines the 2-halo and 1-halo term respectively. Finally, the blue dotted line correspond to the theoretical best fit obtained from the convergence stacking results using a NFW mass profile with the mass and the concentration as free parameters (see Section \ref{sec:constrains} for more details).}
\label{fig:results_xcoQ}
\end{figure}

\subsection{The mass of a cluster size halo around the QSOs acting as lenses}
\label{sec:halomass}
As shown by the red circles in Fig. \ref{fig:results_xcoQ}, we computed the CCF on the QSOs sample introduced in Section \ref{sec:data}, using eq. \ref{eq:xcor} as explained in Section \ref{sec:xcorr}. The comparison with the results obtained on the GN17 sample is also shown (grey squares). Such sample consisted in elliptical galaxies with redshift in the $0.2-0.8$ range. There is general agreement between our findings and those by GN17. We also plot the cross-correlation for the objects in their highest bin of redshift when performing tomographic analysis ($0.5-0.8$, dark grey diamonds), since this is the closest case to the redshift distribution of our QSOs sample, showing that the trend at higher redshift is to stay slightly higher with respect to the CCF of the whole sample.

On the one hand, it should be noticed that the QSOs CCF measurements cover a large angular scale range, from arcsec to few degrees. On the other hand, the relatively small sample translates in big error-bars due to the lack of statistic.

On such data, we perform a halo modelling analysis following exactly the same methodology as in GN17. The CCF between a foreground and background source populations can be expressed as (see GN17 and \cite{Coo02} for more details):
\begin{equation}
w_{fb}(\theta)=\langle\delta N_f(\hat{n})\delta N_b(\hat{n}+\theta)\rangle ,
\end{equation}
being $\delta N_f(\hat{n})$ and $\delta N_b(\hat{n})$ the fluctuation in the number of the foreground and background source populations, respectively. In the presence of lensing, the background sources number counts are modified due to the amplification and the dilution effects (see Section \ref{sec:theory} for a more detailed discussion). For the weak lensing regime the amplification can be approximated by $\mu\simeq 1+2\kappa$ and, therefore, the correlation between the foreground and background sources can be evaluated as (using the standard Limber \cite{Lim53} and flat-sky approximations):
\begin{equation}
w_{fb}(\theta)=2(\beta-1)\int_0^{z_s} \frac{dz}{\chi^2(z)}\, \frac{dN_f}{dz} \mathrm{W^{lens}}(z) \int_0^\infty{\frac{\ell d\ell}{2\pi}P_\mathrm{gal-dm}(\ell/\chi(z),z)J_0(\ell\theta),}
\label{eq:xcor_gen}
\end{equation}
where $P_\mathrm{gal-dm}$ is the cross-correlation power spectrum between the galaxy and dark matter distributions (see GN17 and \cite{Coo02} about how to parametrise it under the halo model formalism), $\frac{dN_f}{dz}$ is the \textit{unit-normalized} foreground redshift distribution and
\begin{equation}
\mathrm{W^{lens}}(z)= \frac{3}{2}\frac{H^2_0}{c^2} E^2(z) \int_z^{z_s}dz'\frac{\chi(z)\chi(z'-z)}{\chi(z')}\frac{dN_b}{dz'}
\label{eq:len_kernel}
\end{equation}
with $\chi(z)$ as the comoving distance to redshift $z$, $E(z)=\sqrt{\Omega_M(1+z)^3+\Omega_\Lambda}$ and $\frac{dN_b}{dz}$ as the \textit{unit-normalized} background redshift distribution. Following \cite{Bia15,Bia16, GN17} we adopted $\beta = 3$ as our fiducial value. 

Normally, the galaxy mean occupation function is represented as the sum of its physically illustrative central and satellite components, $\langle N_{gal}(M)\rangle = \langle N_{cen}(M)\rangle+\langle N_{sat}(M)\rangle$ (e.g. \cite{Zeh04}). 
One conclusion from GN17 is that to produce such a high magnification bias, the halos acting as lenses have a high minimum mass. In our case, this observational threshold is also confirmed by the limited range of the estimated r-band luminosities of the QSOs acting as lenses: $29\lesssim log_{10}(L_r [erg/s/Hz])\lesssim 30$. Therefore we can safely apply the same functional form to describe our sub-sample of QSOs acting as lenses.

In most recent works, the mean occupation function is given as a softened step function for the central component in addition of a power law for the satellite component (e.g. \cite{Ric12,Ric13,Mit18}). In our case, taking into account the big measurement uncertainties and in order to compare directly our results with those of GN17, we adopted a more simple mean occupation function with less free parameters: a step function plus a power law for the satellite component (e.g. \cite{Zhe04}). An halo host a galaxy at its centre when $M>M_{min}$ and the number of satellites is regulated as $N_{sat}(M)=\left( M/M_1\right)^\alpha$, with $M_1$ the pivotal mass when the halos start to host additional satellites and $\alpha$, the power-law exponent. As in \cite{Mit18}, we assumed that the mean occupation function has a negligible redshift evolution, although the foreground and background redshift distributions are considered in the theoretical estimations (see equations \ref{eq:xcor_gen} and \ref{eq:len_kernel}).

Once the cosmological parameters and the $\beta$ are fixed, this halo modelling has just three free parameters: $M_{min}$, $M_1$ and $\alpha$. To constrain the best fit values, we use a Markov chain Monte Carlo (MCMC) approach to compare our theoretical model with the measured signal. Considering that, as in GN17, the model does not well-describe the data in the strong lensing regime (below $\sim{30}$ arcsec), we discard these data when performing the model analysis. More details on the weak lensing validity range will be discussed in sec. \ref{sec:constrains}.

To perform the MCMC we use the open source \texttt{emcee} \cite{emcee} software package. It is a stable, well tested Python implementation of the affine-invariant ensemble sampler. We choose uniform priors for the two mass parameters: $12<log_{10}(M_{min}/M_\odot)<14.5$ and $12.5<log_{10}(M_1/M_\odot)<15.5$. For the $\alpha$ parameter, it was chosen a normal prior with mean value 1.5 and standard deviation of 0.05.
For each signal analysis we generated 100 walkers to perform 500 steps each to ensure good statistical sampling and recovery of the posterior distributions.

\begin{figure}
\centering
\includegraphics[width=\linewidth]{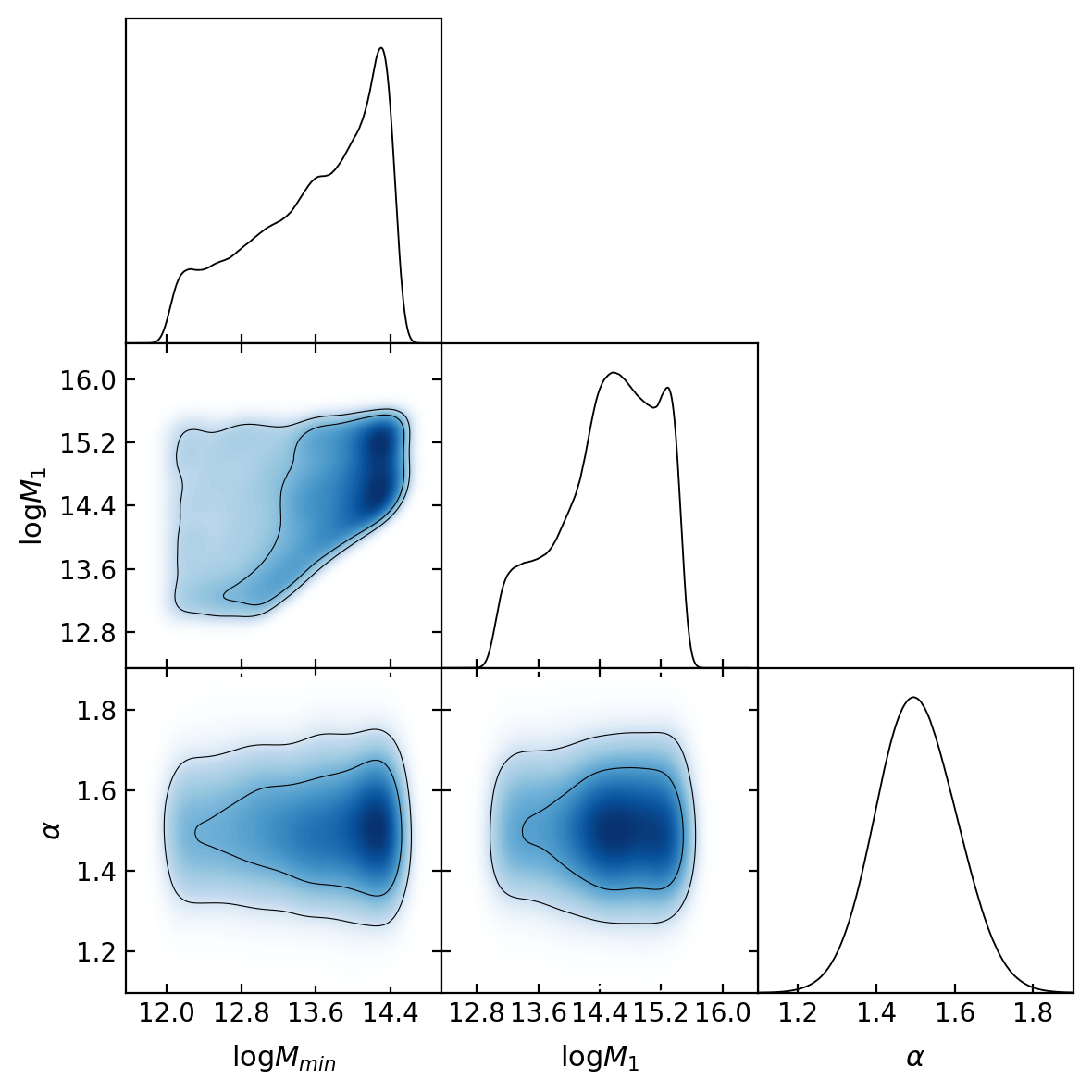}
 \caption{MCMC results summary plot. Estimated CCF MCMC free parameters posterior distributions. The contours levels correspond to 68\% and 95\% of the posterior area.}
\label{fig:tri_plot}
\end{figure}

The results of our MCMC analysis are shown in Fig. \ref{fig:tri_plot}. The contours levels correspond to $68\%$ and $95\%$ of the posterior area. The best fit halo model is shown by the black line in Fig. \ref{fig:results_xcoQ}, being the dotted and dashed lines the 2-halo and 1-halo term respectively. There is a good agreement with the data, at least at angular scales greater than $\gtrsim20$ arcsec. The best fit values are (mean and $68\%$ confidence intervals): $\log_{10}{(M_{min}/M_\odot)} = 13.6_{-0.4}^{+0.9}$ and $\log_{10}(M_{1}/M_{\odot}) = 14.5_{-0.3}^{+0.9}$. 

QSOs selected at different wavelengths have been shown to have typical masses of $0.5 - 10\times10^{12} h^{-1} M_\odot$, almost constant in the full redshift range, $0.5\lesssim z \lesssim 4$, probed by several studies (e.g. \cite{Mye08,Cro05,She07,Pad07,Lea15, Mit18, Ric12,Ric13}). The estimated mass of the QSOs acting as lenses is clearly much higher than these observations. It is simply the usual observational bias in lensing analysis when only the most massive objects from the parent population are able to produce a measurable signal (e.g. \cite{GN14,GN17,Bau14}). Instead of an issue, this result suggests that the QSOs acting as lenses in our sample are not isolated and probably reside in galaxy groups/clusters halos (see below).

It should be noticed that we are aware that the obtained MCMC results are not very satisfactory and this is mostly due to the large uncertainties in our cross-correlation measurements. In particular, the input parameter $\alpha$ is not constrained at all; it depends only on the angular scales for the transition between the 1- and 2-halo regimes. Moreover, the $\log{M_{min}}$ posterior distribution shows the tendency to prefer higher values getting close to the upper prior limits. This is a very well known issue due to the excess of the cross-correlation signal at degree angular scales. It was already found in GN17 in the highest (and most similar to our sample's) redshift bin ($0.5<z<0.8$). What is happening is that the MCMC sampler tends to increase $M_{min}$ in order to fit the 2-halo term at the expanse of exceeding the measured signal at arcmin scales. Given such uncertainties with the current sample, we consider worthless trying a deeper analysis to seek for further, more accurate, results in this part of the analysis.

However, we can use these constraints to discuss about the QSOs host halo by comparing our CCF results obtained with the QSOs acting as lenses against those using ellipticals galaxies as lenses. It should be noticed that GN17 obtained better constraints with respect to the present work since they had a larger lens sample that allow smaller uncertainties.
The minimum halo mass derived for the QSOs acting as lenses is larger than the one derived with the whole sample of Luminous Red Galaxies (LRGs, $0.2<z<0.8$), $\log_{10}{(M_{min}/M_\odot)} = 13.06_{-0.06}^{+0.05}$. This could be interpreted as the QSOs residing in more massive halos when compared with typical elliptical galaxies acting as lenses. But, at the same time, the QSOs halo mass is in good agreement if compared only with the elliptical galaxies in the highest redshift bin analysed ($0.5<z<0.8$) with $\log_{10}{(M_{min}/M_\odot)} = 14.36_{-0.10}^{+0.14}$ (even more if we take into account the maximum of the posterior distribution for the QSOs, $\log_{10}{(M_{min}/M_\odot)} \sim 14.2$). In GN17, the evolution of the halo mass with redshift was interpreted as an observational bias caused by the variation of the lensing probability \cite{Lap12}. We do not have additional evidences to derive different conclusions on the QSOs host halo masses. Finally, considering these results, it seems that, when selected as lenses, both QSOs and LRGs resides in galaxy groups/clusters halos with similar statistical mass properties, i.e. we are observing the lensing effect of a cluster size halo were the QSOs or the LRGs are probably situated in its center with a typical Bright Central Galaxy (BCG) as their galaxy host.

\section{Projected mass density profile}
\label{sec:denprof}
In this section we use the results described in the previous section to estimate the projected mass density profile of the lenses and to derive some constrains using the most common theoretical models.

\subsection{Theoretical background}
\label{sec:theory}
In the presence of gravitational lensing the integrated source number counts per unit area ($n(>S_0)=\frac{dN}{d\Omega dS}(>S_0)=n_0 S_0^{-\beta}$, with $S_0$ the unlensed flux density and $\beta$ the source number count slope) suffer both a dilution and an amplification effects:
\begin{equation}
    n(>S)=\frac{n_0}{\mu(\theta)}\left(\frac{S_0}{\mu(\theta)}\right)^{-\beta}=\mu^{\beta-1} n(>S_0),
\end{equation}
being $\mu$ the amplification factor, defined as the ratio between the lensed to the unlensed flux densities of the source. 
The angular CCF of two population of sources, $w_x(\theta)$, is the fractional excess probability, relative to a random distribution, of finding a source of Population 1 and a source of Population 2, respectively, within infinitesimal solid angles separated by an angle $\theta$ \cite{PEE80}. From the point of view of the lens it can be written as:
\begin{equation}
    n(>S)=n_0(1+w_x(\theta)),
\end{equation}
i.e. the excess of background galaxies around the lens with respect to a random distribution.
Therefore, the CCF can be related with the amplification factor as:
\begin{equation}
    w_x(\theta)=\frac{n}{n_0}-1=\mu(\theta)^{\beta-1}-1.
    \label{eq:xcorr_mu}
\end{equation}
The magnification can be expressed as a function of the \textit{convergence}, $\kappa$ and \textit{shear}, $\gamma$, as 
\begin{equation}
    \mu=\frac{1}{(1-\kappa)^2+|\gamma|^2}.
    \label{eq:amplification}
\end{equation}
For unresolved/point like background sources the shear can be considered negligible at all angular scales, as all the galaxies have the same point spread function. In other words, the excess of detection probability that is producing the magnification bias is not affected by the potential small shape distortions produced by the shear (that in any case is completely negligible in the weak lensing regime). As a consequence, we can estimate the convergence as a function of the CCF as:
\begin{equation}
    \kappa(\theta)=1-(w_x(\theta)+1)^\frac{-1}{2(\beta-1)}.
    \label{eq:kappa}
\end{equation}
Notice that this expression is exact and we did not perform any approximation to obtain it. In fact, it can be validly applied in all lensing regimes.
A more simple relationship can be found if we assume the weak lensing approximation ($\kappa\longrightarrow 0$ and, therefore, $\mu\sim1+2\kappa$):
\begin{equation}
    \kappa(\theta)\simeq \frac{w_x(\theta)}{2(\beta-1)}.
    \label{eq:kappa_WL}
\end{equation}
Finally, the surface mass density, $\Sigma(\theta)$, is defined as $\kappa(\theta)=\frac{\Sigma(\theta)}{\Sigma_{crit}}$, being $\Sigma_{crit}=\frac{c^2}{4\pi G}\frac{Ds}{D_dD_{ds}}$ the critical surface density with $D_d$ and $D_s$ the proper diameter distance from the observer to the deflector and source, respectively. $D_{ds}$ is the proper diameter distance from the deflector to the source. In general, since we are dealing with samples of objects (lenses and background sources) we need to use their redshift averaged distances (for example as in the cross-correlation case):
\begin{equation}
    \left< D_{pop} \right>=\left[ \int_0^\infty dzN_{pop}(z)D_{pop}(z)\right]\left[ \int_0^\infty dz N_{pop}(z)\right]^{-1},
\end{equation}
with $N_{pop}(z)$ the redshift distribution function for each sample. However, in the stacking case we have also the information about the redshift for each individual QG pair that allow us to estimate $\Sigma_{crit}$ individually for each QG pair and then calculate the average.

\subsection{Mass density profiles}
\label{sec:profiles}
We assume that the mass of the lenses is dominated by dark matter and, therefore we model the dark matter halo as a Navarro-Frenk-White (NFW) profile \cite{NFW96}:
\begin{equation}
    \rho_{NFW}(r)=\frac{\delta_c \rho_{crit}(z)}{(r/r_s)[1 + r/r_s]^2},
\end{equation}
where $\delta_c=\frac{200}{3} \frac{C^3}{ln(1+C)-C/(1+C)}$, with $C$ the concentration parameter. The \textit{virial radius} is denoted by $r_{200c}$ and is the radius inside which the mean mass density of the halo equals $200 \rho_{crit}(z)$. The total mass inside $r_{200c}$ is $M_{200c}$ with $r_s=r_{200c}/C$ and $\rho_{crit}(z)$ is the critical density of the universe at the redshift of the cluster.

To take into account the variation of the concentration of haloes with mass and redshift, we assume the relation measured in \cite{Man08}:
\begin{equation}
    C(M_{200c}, z) = \frac{4.6}{1+z}\left( \frac{M_{200c}}{1.56\times 10^{14} h^{-1} M\odot}\right)^{-0.13}
\end{equation}
Then the convergence can be estimated as a projection along the line-of-sight (\cite{Bar96,Sch05}):
\begin{equation}
    \kappa_{NFW}(\theta)=\frac{2r_s\delta_c\rho_{crit}(z)}{\Sigma_{crit}} f(\theta/\theta_s),
\end{equation}
with $\theta_s=r_s / D_d$, the angular scale radius, and
\begin{equation}
    f(x)=\frac{1}{x^2-1}[1-\mathcal{F}(x)],
\end{equation}
where
\begin{equation}
    \mathcal{F}(x)=
    \begin{cases}
       \frac{acosh(1/x)}{\sqrt{1-x^2}} & \text{for $x<1$} \\
       1/3 & \text{for $x=1$} \\
       \frac{acos(1/x)}{\sqrt{x^2-1}} & \text{for $x>1$}        
    \end{cases}
\end{equation}

We are considering only the 1-halo term. The 2-halo was already studied directly with the CCF measurements and the halo modelling analysis in section \ref{sec:xcorr}. In this case the only free parameter of the model is the mass of the halo, $M_{200c}$.

This profile will be compared with the other commonly used model, the classical singular isothermal sphere density profile (SIS):
\begin{equation}
    \rho_{SIS}(r)= \frac{\sigma_{SIS}^2}{2\pi G r^2},
\end{equation}
where $\sigma_{SIS}$ is the one-dimensional velocity dispersion of the mass that is commonly assumed as $\sigma_{SIS}\simeq V_H/\sqrt{2}$, with the halo circular velocity $V^2_H=G M_H/r_H$.

Similarly to the NFW profile, the convergence can be estimated as a projection along the line-of-sight \cite{Sch05}:
\begin{equation}
    \kappa(\theta)=\frac{\theta_E}{2|\theta|},
\end{equation}
with $\theta_E=4\pi\left( \frac{\sigma_{SIS}}{c}\right)^2 \frac{D_{ds}}{D_s}$, the Einstein radius in this lens model. As in the NFW profile, the only free parameter is the halo mass $M_{200}$.

\subsection{Results and Constrains}
\label{sec:constrains}

\begin{figure*}
\centering
\includegraphics[width=0.8\linewidth]{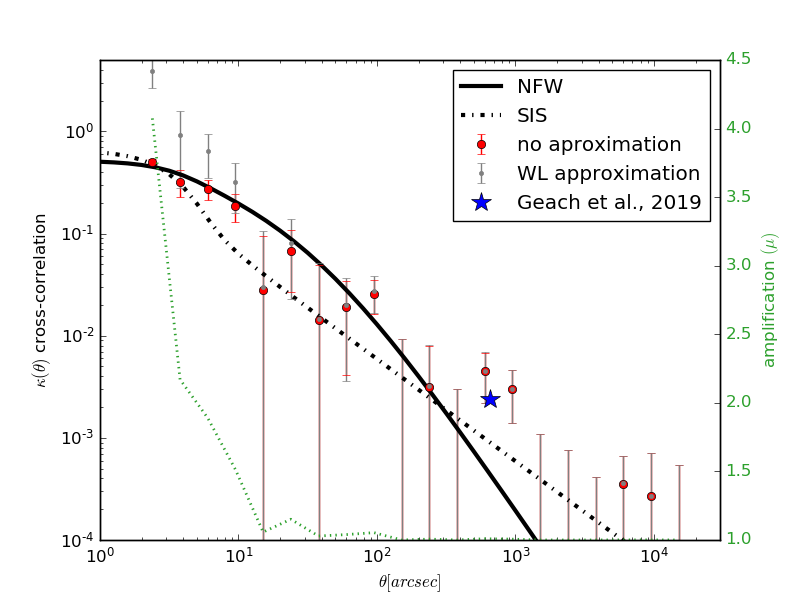} \\
\includegraphics[width=0.8\linewidth]{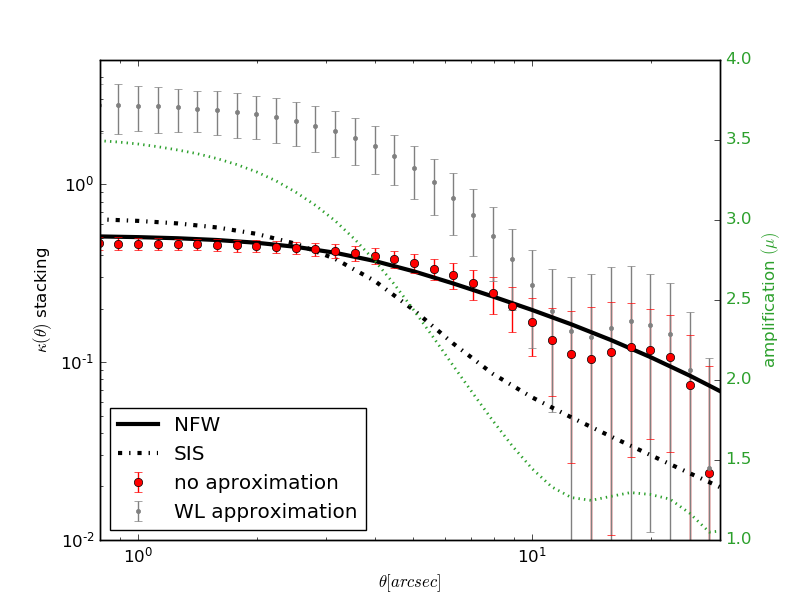}
\caption{Estimated convergence, $\kappa(\theta)$, using the traditional CCF estimator (top panel) and the stacking approach data (bottom panel), with (red circles) and without (grey stars) the weak lensing approximation. The green dotted line is the amplification, whose scale is shown on the right y-axis. The blue star is the the value obtained by \cite{GEA19} by staking SDSS QSOs on \textit{Planck} estimated convergence map. In both panels the best-fit using different mass density profiles are also shown: NFW with mass and C as free parameters(solid line) and the SIS (dot-dashed line).}
 \label{fig:convergence}
\end{figure*}

\begin{figure*}
\centering
\includegraphics[width=0.8\linewidth]{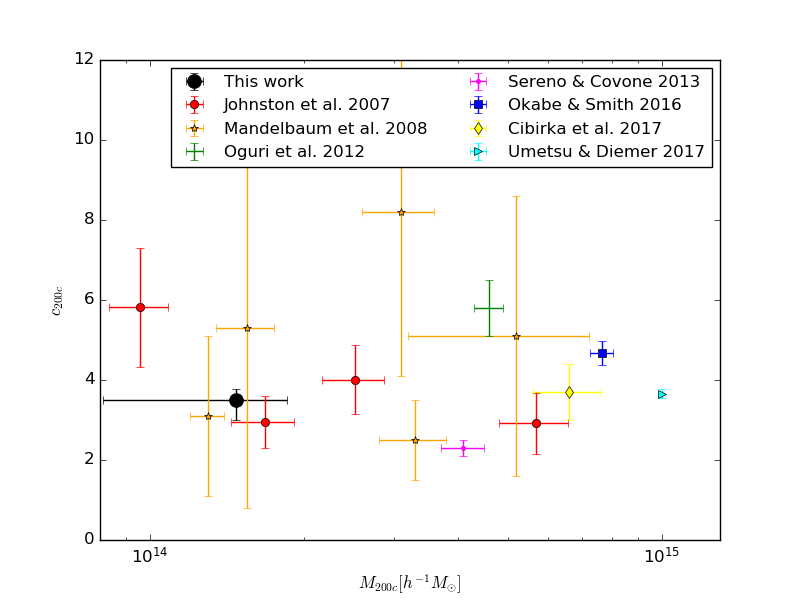}
\caption{Concentration estimated using our findings (big black dot) compared with concentration values from previous work \citep{Joh07, Man08, Ogu12, Ser13, Oka16, Cib17, Ume17}.}
 \label{fig:concentration}
\end{figure*}

From eq. \ref{eq:kappa}, it is clear that there is a simple transformation between the CCF and the convergence. Therefore, both of them are equally valid quantities to be used to study the mass density profile of the associated lens halos. However, there are some points to be taken into account before choosing between the two quantities:
\begin{itemize}
    \item The CCF is a very specific measurement that is difficult to be compared with other lensing or mass density analyses. On the other hand, the convergence, or the directly related surface mass density, is a more common quantity usually estimated by different techniques (e.g \cite{Ume14,Ume16}) that can be used to make comparisons between different probes.
    \item By assuming a particular mass density profile, it is possible to estimate the 1-halo term that describe the CCF at angular scales $\lesssim 1$ deg for both the strong and weak lensing regimes. However, it is still needed a 2-halo term to fully characterise the full angular scale range of the measured CCF (as described in sec. \ref{sec:xcorr}).
    \item Different theoretical approaches provide different information. The mass density profile, for example in the case of a NFW profile, only depends on two parameters, total mass and concentration, but can only explain the 1-halo term. To describe the larger scales an additional 2-halo term is required, that can be almost studied independently of the 1-halo term. It can be approximately expressed as a function of the halo mass and the lens bias (e.g. \cite{Bau14}). On the contrary, if we assume an HOD approach (as e.g. in GN17 based on \cite{Coo02}) we can obtain information on a different set of parameters as $M_{min}, M_1, \alpha$ and bias. The same theoretical framework describes both the 1-halo and 2-halo terms, but it assumes the weak lensing approximation and cannot be applied in the strong lensing regime ($\lesssim 20$ arcsec).
    \item Differently from the halo modelling of the auto-correlation function, the 2-halo term of the CCF depends also on $M_1$ and $\alpha$ that are mostly related with what happen inside the single halo or 1-halo term. Therefore, if we try to describe the 1-halo term only with the mass density profile approach, the 2-halo term will be described by too many parameters not directly related with the remaining data (angular scale measurements above 1 deg).
\end{itemize}
The conclusion is that no single approach seems to be able to describe the full range of CCF measurements: the halo modelling assumes the weak lensing approximation, so $\theta\gtrsim 20$ arcesc, while the mass density profile approach can be applied only to describe the 1-halo term, so $\theta\lesssim1$ deg.
For these reasons, in this work we have preferred to perform the traditional halo modelling analysis in the weak lensing regime and to use the convergence to investigate the mass density profile and restrict the analysis to the 1-halo term angular scales.

Figure \ref{fig:convergence} shows the convergence, $\kappa(\theta)$, computed using traditional CCF measurements (top panel, Table \ref{tab:k-xcorr}) and the new stacking approach (bottom panel, Table \ref{tab:k-stacking}). Mind the fact that the angular scales of the two plots are different: as described before, the traditional CCF derived results are preferred at large angular scales, whereas the stacking ones are better-performing at smaller scales.
Moreover, it has to be noticed that the stacking CCF measurements are estimated from a smoothed map to take into account the positional uncertainty of the H-ATLAS background sample. It was used a 2.4 arcsec as the standard deviation. The same smoothing was applied to the theoretical convergence, estimated for the different mass density profiles, before performing any comparison.

The blue star in the top panel shows the perfect agreement with the value obtained by \cite{GEA19} by stacking nearly 200,000 targets on \textit{Planck} estimated lensing convergence map. The targets are QSOs from the SDSS, covering the redshift range $0.9 \leq z \leq 2.2$.


As can be seen in the top panels of Fig. \ref{fig:analysis_4Z} and already mentioned in the description of Fig. \ref{fig:results_npix100}, we do not have enough QG pairs to produce an homogeneous map at angular separation above several arcsec. This fact is translated into a noisy estimated convergence above 15 arcsec in the stacking case, and even worse in the cross-correlation one. With larger samples we should be able to estimate a smoother convergence at larger angular separations.

In both cases, we compared the results from the convergence given by the formulas eq. \ref{eq:kappa} and \ref{eq:kappa_WL}. It should be noticed that the weak lensing approximation overestimates the general formula results whenever the convergence value is greater than $\sim 0.1$, independently of the angular scale. 
Given the range of masses of our lenses this result implies that both formulas agree at large scales, but start to differ below 20-30 arcsec. The angular scale range for the transition between the strong and weak lensing regimes is in agreement with the findings by GN17 when analysing the halo model fit results. Moreover, it is also confirmed by our own halo model fit results in section \ref{sec:halomass}.

In addition, by calculating the amplification factor (eq. \ref{eq:amplification}) we also showed that these angular scales indicate the transition toward a strong lensing regime with amplification factors rapidly increasing well above 1.5 at angular scales smaller than 20 arcsec (see Fig. \ref{fig:convergence}, green dotted line). The importance of this conclusion is in the fact that many lensing studies rely on the weak lensing approximation, pushing it to the smallest possible scales, where such approximation is no longer valid. In fact, it is very clear from the stacking approach results (Fig. \ref{fig:convergence}, bottom panel) that at small scales both estimations differ by almost an order of magnitude.

Our halo convergence measurements using QSOs as lenses are in good agreement with previous estimations, even if some of them used the magnification bias with lenses of different kinds (\cite{Bau14}) or first performed several analysis of the strong and weak lensing in individual clusters and then stacked them together (see Fig. 7 in \cite{Ume14} and Fig. 3 in \cite{Ume16}). In both cases there is an intrinsic limitation to the minimum physical size that can be studied due to the presence of the lens itself. 
In the first case, the lensing mass profiles of spectroscopic LRGs and galaxy clusters were determined through measurements of the weak lensing magnification of photometric LRGs in their background (the change in detected galaxy counts as well as the increased average galaxy flux behind the lenses). The fact that the background sample is in the optical band as the lenses implies that the LRGs or the BCG complicates the measurements at angular scales smaller than the lens size.
In the second case, the strong lensing information is provided by the analysis of arcs or multiple images of the same object. But when they happen at small angular scales, such arcs or images cannot be detected near the BCGs. For these reasons the mass profile reconstructions usually reach only several kpc.

In our case, the fact that the foreground and background samples were observed in different wavelengths bands (in particular, the sub-mm emission of the QSOs and the optical emission of the SMGs are negligible) allow us to overcome this problem, achieving measurements below 10 kpc (considering a scale factor of $\sim7.2$ kpc/arcsec at the lenses mean redshift). This limit is only imposed by the samples' statistics and the positional uncertainty of the instruments. As discussed below, even taken into account the positional uncertainties it is possible to derive interesting conclusions at such small scales.

Using the mass density profiles described above, we estimate their main parameters that provide the best fit to the data. We apply the same smoothing scale to the theoretical density profiles before the comparison. It did not introduce any additional uncertainty to the comparison because we used exactly the same standard deviation in both cases. Moreover, we use the estimated covariance matrix to calculate the chi-squared during the best fit analysis. For the NFW profile, we performed the best-fit only to the data obtained with the stacking CCF. Even after the smoothing, it is clear that the SIS model cannot provide a good fit to the stacked data (the higher slope produces a very distinctive enhancement around the smoothing scale that is difficult to be adjusted to the measurements). Therefore, we estimated the SIS $M_{200c}$ parameter using only the traditional CCF measurements.

The NFW profile provided a good fit to the estimated convergence at angular scales below $\sim 1$ arcmin ($\sim400$ kpc). This result is an additional confirmation of the cuspy nature of the dark matter halos expected from dark matter only simulations \cite{Wec02}. Beyond 200 arcsec ($\sim$ 2 Mpc) the NFW profile would require a 2-halo term in order to fit the measurements. As we are interested here in the inner halo profile, and the LSS structure was already studied in section \ref{sec:xcorr}, we decided not to calculate the NFW 2-halo term.

With the NFW profile we estimate a $M_{200c}=1.7_{-0.5}^{+2.1}\times10^{14} M_\odot$. The uncertainties are dominated mainly by the redshift distribution of both samples, while the parameter estimation errors are negligible in comparison. This estimation was obtained using the individual QG pair redshift information: the $\Sigma_{crit}$ was estimated for each QG pair and the average and dispersion of all the QG pairs was used. There is no relevant difference if we had used the overall redshift distribution of the parent samples (a single $\Sigma_{crit}$ value estimated at the mean redshifts): $M_{200c}=1.4_{-1.0}^{+2.2}\times10^{14} M_\odot$. As seen in Fig. \ref{fig:pairs_dndz}, the redshift distribution of the QG pairs are representative of the parent samples ones.

As a comparison, with the SIS profile we obtain $M_{200c}= 1.6_{-0.4}^{+0.7}\times10^{13} M_\odot$. It is a lower value with respect to the one for the NFW case: it is mainly due to the fact that the SIS does not provide a good fit to the data below 20 arcsec and that some of the traditional CCF measurements have low values with large uncertainties.

The estimated masses from the profile analysis support the results and conclusions derived through the halo modelling performed before (see sec. \ref{sec:halomass}).
In fact, the estimated mass is in agreement with the ones derived by \cite{Bau14} both for the LRGs and the galaxy clusters (almost independent of the richness of the clusters) based also in the magnification bias. Although not discussed in that work, the fact that they derived similar masses for the LRG sample and the cluster ones indicates, as in \cite{GN17} and in our case, that the LRGs were sign-posting cluster size halos. It seems that in all these works the measured magnification bias is dominated by the cluster size halo effect, being the galactic size halos contribution almost negligible.

On this respect, we tried to detect any residual baryonic/stellar component at very small scales by adding a second independent SIS profile. Due to the particular shape of the SIS profile, the fit to the data worsened for any $M_{200c}\gtrsim 10^{12} M_\odot$. For lower masses the potential contribution of any baryonic/stellar component became completely negligible.

However, the study by \cite{GEA19} indicates that quasars with i-band magnitudes, $M_i$  brighter than $M_i \lesssim -26$ reside in halos of typical mass $M_h \sim 10^{13} h^{-1} M_\odot$. This mass value is in agreement with the typical host halo QSO masses derived with other techniques as discussed earlier and similar to our mass estimation using the SIS profile at large scales. There are two main reasons that explain the difference with respect to our main findings. On the one hand, they are stacking on a convergence map. This means that all the targets contribute to the final measurement even in a small fraction due to their smaller mass. In the case of the magnification bias we can only count if there is a pair or not, and therefore only the effect of the more massive lenses is observed. On the other hand, they use a more than 100 times larger target sample that allows them to be sensitive to the effect of much weaker, i.e. less massive, objects. Although their results are more representative of the bulk of the QSO host population, they are limited to a very narrow angular separation regime that restrains the precise analysis and applications of their measurements: only angular scales above the \textit{Planck} resolution can be studied and the measurements above a few Planck beams are compatible with zero. On the contrary, the magnification bias study has a clear observational bias towards higher lens masses but can be robustly measured and analysed for a wide range of angular scales, from arcsec to degrees.

The \cite{Man08} mass-concentration relationship was chosen because it was obtained from stacked weak lensing measurements. However, when compared with more recent relations, it is shown to provide a rather low concentration values. As a test, we estimated the halo mass using two different, most recent, relations: \cite{Dut14} and \cite{Chi18}. We found $M_{200c}=0.8_{-0.1}^{+0.5}\times10^{14} M_\odot$ and $M_{200c}=0.9_{-0.2}^{+0.6}\times10^{14} M_\odot$, respectively. The higher concentration values produce lower mass, but, in any case, both estimations are compatible at $1\sigma$ level with the previous one using \cite{Man08}.

If the concentration parameter is considered also as a free parameter, the best-fit NFW profile is obtained for the following parameter values: $M_{200c}=1.0_{-0.2}^{+0.4}\times10^{14} M_\odot$ and $C=3.5_{-0.3}^{+0.5}$ (solid line in both panels of Fig. \ref{fig:convergence}). This result is in good agreement with the \cite{Man08} relationship, as well as with other stacked results as \cite{Joh07}, $z\sim 0.15$ and \cite{Ser13} at $z\sim 1$. However, at this mass range and redshift, these results are below typical relations as the \cite{Dut14} and \cite{Chi18} ones, that consider mainly relaxed halos.

Halos are dynamically evolving objects and therefore their mass and concentration is probably related with their recent assembly history \cite{Ser13}. In fact, after a recent merger the halo profile may not be well-described by the NFW profile due to a dynamically unrelaxed state \cite{Chi18}, although this problem should be mitigated in stacked halo profiles. Following \cite{Chi18} simulation results, in unrelaxed halos much of the mass is far from the center and therefore they tend to have lower concentrations than the relaxed ones. Thus, our derived mass-concentration results could be interpreted as the QSOs acting as lenses are mostly hosted in unrelaxed cluster size halos. It could be related with the observed link between QSO activity and merger episodes at low redshift (see e.g. \cite{Lap06} or \cite{Som15}). In any case, using higher concentration values also produced a compatible fit to our observations within the error bars. More QG pairs with angular separation around 20-30 arcsec are needed to obtain smaller error bars in order to negate concentration values higher than 4 for our QSOs sub-sample.

Finally, by considering the best fit NFW profile based on the stacking CCF results with the halo mass and the concentration as free parameters, we used equations \ref{eq:amplification} and \ref{eq:xcorr_mu} to estimate the corresponding CCF (dotted blue line in Fig. \ref{fig:results_xcoQ}). Taking into account that it was estimated using the stacking results with data limited to 40--50 arcsec, it provides a valid fit to the observed CCF below $\sim 1$ deg. However, when compared with the halo modelling results, it is clear that the 1-halo term from the halo modelling approach is below the other one. This is mainly due to the fact that the weak lensing approximation needed to derive the halo modelling for the traditional CCF implies that only data above $\sim 20$ arcsec can be used. Those measurements have large uncertainties and lower values compared with other more accurate data points at other angular distances (e.g. those in the strong lensing regime). With more accurate observations, as the ones for elliptical galaxies from GN17, both approaches should provide similar results.

In any case, for future works on this topic, it would be interesting to combine both approaches in a 2-steps analysis: first, characterise the mass density profile for the particular sample by using the stacking CCF results and then use this information in a traditional halo modelling to describe the CCF at larger scales.

\begin{table}[tbp]
    \centering
    \begin{tabular}{|c|c|c||c|c|c|}
    \hline
    $\theta$ [arcmin] & $\kappa$ & $err_{\kappa}$ & $\theta$ [arcmin] & $\kappa$ & $err_{\kappa}$\\
    \hline
    0.04 & 0.50 & 0.04 & 3.983 & 0.003 & 0.005\\  
    0.064 & 0.32 & 0.09 & 6.31 & $< 0.003$ & -\\
    0.10 & 0.27 & 0.06 & 10.00 & 0.004 & 0.002\\
    0.16 & 0.19 & 0.06 & 15.85 & 0.003 & 0.002\\
    0.25 & 0.03 & 0.07 & 25.12 & $< 0.001$ & -\\
    0.40 & 0.07 & 0.04 & 39.81 & $< 0.001$ & -\\
    0.63 & 0.01 & 0.03 & 63.10 & $<  0.0004$ & -\\
    1.00 & 0.02 & 0.02 & 100.00 & 0.0004 & 0.0003\\
    1.58 & 0.03 & 0.01 & 158.49 & 0.0003 & 0.0004\\
    2.51 & $ < 0.01$ & - & 251.19 & $< 0.0005$ & -\\
    \hline
    \end{tabular}
    \caption{\label{tab:k-xcorr} Angular distance, convergence and its error computed using the traditional CCF data.}
\end{table}
\begin{table}[tbp]
    \centering
    \begin{tabular}{|c|c|c||c|c|c|}
    \hline
    $\theta$ [arcsec] & $\kappa$ & $err_{\kappa}$ & $\theta$ [arcsec] & $\kappa$ & $err_{\kappa}$ \\
    \hline
    0.10 & 0.50 & 0.2 & 3.16 & 0.44 & 0.03\\
    0.11 & 0.50 & 0.09 & 3.55 & 0.43 & 0.03\\
    0.16 & 0.50 & 0.09 & 3.98 & 0.41 & 0.03\\
    0.22 & 0.50 & 0.04 & 4.47 & 0.39 & 0.03\\
    0.32 & 0.50 & 0.04 & 5.01 & 0.37 & 0.04\\
    0.35 & 0.50 & 0.06 & 5.62 & 0.35 & 0.04\\
    0.40 & 0.50 & 0.06 & 6.31 & 0.33 & 0.04\\
    0.45 & 0.50 & 0.04 & 7.08 & 0.31 & 0.04\\
    0.50 & 0.50 & 0.04 & 7.94 & 0.28 & 0.04\\
    0.56 & 0.50 & 0.05 & 8.91 & 0.25 & 0.05\\
    0.63 & 0.50 & 0.03 & 10.00 & 0.22 & 0.06\\
    0.71 & 0.50 & 0.03 & 11.22 & 0.17 & 0.07\\
    0.79 & 0.50 & 0.04 & 12.59 & 0.13 & 0.09\\
    0.89 & 0.50 & 0.04 & 14.13 & 0.1 & 0.1\\
    1.00 & 0.50 & 0.03 & 15.85 & 0.1 & 0.1\\
    1.12 & 0.49 & 0.03 & 17.78 & 0.13 & 0.09\\
    1.26 & 0.49 & 0.03 & 19.95 & 0.15 & 0.07\\
    1.41 & 0.49 & 0.03 & 22.39 & 0.15 & 0.07\\
    1.58 & 0.49 & 0.03 & 25.12 & 0.10 & 0.07\\
    1.78 & 0.48 & 0.03 & 28.18 & 0.08 & 0.07\\
    2.00 & 0.48 & 0.03 & 31.62 & 0.11 & 0.07\\
    2.24 & 0.47 & 0.03 & 35.48 & 0.10 & 0.07\\
    2.51 & 0.46 & 0.03 & 39.81 & 0.12 & 0.06\\
    2.82 & 0.45 & 0.03 & 44.67 & 0.08 & 0.06\\
    \hline
    \end{tabular}
    \caption{\label{tab:k-stacking}Angular distance, convergence and its error computed using the stacking CCF data.}
\end{table}

\section{Conclusions}
\label{sec:concl} 
In this work we were able to measured the gravitational lensing effect (magnification bias) produce by QSOs acting as lenses, $0.2<z<0.8$, on the SMGs observed by Herschel at $1.2<z<4.0$. Although in literature there are at least four confirmed detections of individual QSOs acting as lenses, this is actually the first time that some statistical constrains about their halo mass or density profile can be derived exploiting the lensing effect.

The CCF between the two samples were measured from $\sim5$ arcsec up to $\sim3$ degrees. By performing a halo modelling, the estimated best fit values were (mean and $68\%$ confidence intervals): $\log_{10}{(M_{min}/M_\odot)} = 13.6_{-0.4}^{+0.9}$ and $\log_{10}(M_{1}/M_{\odot}) = 14.5_{-0.3}^{+0.9}$. As in \cite{GN17}, these mass values indicate that the QSOs acting as lenses are placed in galaxy groups/clusters halos, i.e. we are observing the lensing effect of a cluster size halo were the QSOs are probably situated at its centre. Although the SMGs unique properties allow us to obtain the CCF measurements, the lack of a better statistics affects their signal-to-noise and therefore the accuracy of our derived constraints.

We also proposed the stacking technique as an alternative or at least a complementary methodology to study the magnification bias. It allowed us to obtain an averaged QG pairs distribution map around the lenses, that can then be used to estimate the radial CCF. This alternative CCF estimation is in good agreement with the results from the traditional approach, confirming that they are both describing the same physical quantity. The stacking technique provides a way to take into account the positional uncertainty of the data, otherwise not possible with the traditional CCF estimators, that allow us to explore scales of the order of $\sim 1$ arcsec. For this reason, it is the preferable methodology to study the magnification bias in the strong lensing regime, while, at larger scales, it would be more suitable and straightforward to use the traditional approach. Moreover, the stacking technique allow us to know the QG pair contributing to the magnification bias measurements that helps to derive more accurate constraints than in the traditional CCF case.

The lensing convergence, $\kappa(\theta)$, was derived from the estimated CCF measurements. The weak lensing approximation was compared with the exact formula (neglecting the shear contribution, reasonable in our case). Our results showed that, below $\sim 20-30$ arsec, the two formulas start to differ from each other, which is well explained by the fast increase of the estimated amplification above 1.5. Therefore, at about such angular scale the transition between the weak and strong lensing regimes happens. It confirms what has been previously concluded in GN17 and again in the current work by the halo modelling analysis. It should be noticed that applying the weak lensing approximation also in the strong lensing regime would have completely changed our conclusions on the mass density profile.

As expected for cluster size halos, the estimated convergence of the QSOs cluster size host halo is well represented by a NFW mass density profile. The best fit mass was found to be of $M_{200c}\sim 1\times10^{14} M_\odot$, almost independent of the particular mass-concentration relation used. No additional baryonic/stellar component was detected at the smallest scales.

 Finally, by estimating also the concentration parameter we obtained $M_{NFW}=1.0_{-0.2}^{+0.4}\times10^{14} M_\odot$ and $C=3.5_{-0.3}^{+0.5}$. This concentration value is rather low compared with other estimations at similar mass range and it would indicate that the cluster size halos around these QSOs are unrelaxed (following \cite{Chi18} simulation results). However, higher concentration values still provide a compatible fit to the data.

\acknowledgments
We very much thank the reviewer for the valuable comments that helped us to improve this work.
\\
LB and JGN acknowledge the PGC 2018 project PGC2018-101948-B-I00 (MICINN/FEDER) and PAPI-19-EMERG-11 (Universidad de Oviedo). LB, JGN and SLSG acknowledge financial support from the I+D 2015 project AYA2015-65887-P (MINECO/FEDER). JGN acknowledges financial from the Spanish MINECO for a "Ramon y Cajal" fellowship (RYC-2013-13256).
FB acknowledges support from an Australian Research Council Future Fellowship (FT150100074). AL acknowledges partial support from PRIN MIUR 2015 Cosmology and Fundamental Physics: illuminating the Dark Universe with Euclid, by the RADIOFOREGROUNDS grant (COMPET-05-2015, agreement number 687312) of the European Union Horizon 2020 research and innovation program, and the MIUR grant 'Finanziamento annuale individuale attivita base di ricerca'. MN received funding from the European Union's Horizon 2020 research and innovation programme under the Marie Sklodowska-Curie grant agreement No 707601. Finally, EDA and JDCJ acknowledge financial support from the I+D 2017 project AYA2017-89121-P and JDCJ acknowledges support from the European Union's Horizon 2020 research and innovation programme under the H2020-INFRAIA-2018-2020 grant agreement No 210489629.
\\
The analysis were partly carried out at the `Centro Interuniversitario del Nord-Est per il Calcolo Elettronico' (CINECA, Bologna), with CPU time assigned under the project Sis18\_COSMOGAL, ref. tts\#345223.
\\
The Herschel-ATLAS is a project with Herschel, which is an ESA space observatory with science instruments provided by European-led Principal Investigator consortia and with im- portant participation from NASA. The H-ATLAS website is http://www.h-atlas.org/
\\
Funding for SDSS-III has been provided by the Alfred P. Sloan Foundation, the Participating Institutions, the National Science Foundation, and the U.S. Department of Energy Office of Science. The SDSS-III web site is \url{http://www.sdss3.org/}. SDSS-III is managed by the Astrophysical Research Consortium for the Participating Institutions of the SDSS-III Collaboration including the University of Arizona, the Brazilian Participation Group, Brookhaven National Laboratory, Carnegie Mellon University, University of Florida, the French Participation Group, the German Participation Group, Harvard University, the Instituto de Astrofisica de Canarias, the Michigan State/Notre Dame/JINA Participation Group, Johns Hopkins University, Lawrence Berkeley National Laboratory, Max \textit{Planck} Institute for Astrophysics, Max \textit{Planck} Institute for Extraterrestrial Physics, New Mexico State University, New York University, Ohio State University, Pennsylvania State University, University of Portsmouth, Princeton University, the Spanish Participation Group, University of Tokyo, University of Utah, Vanderbilt University, University of Virginia, University of Washington, and Yale University.
\\
This research has made use of the Ned Wright's Cosmology Calculator \cite{nedcc}, and the python packages \texttt{emcee} \cite{emcee}, \texttt{ipython} \cite{ipython}, \texttt{matplotlib} \cite{matplotlib}, \texttt{Scipy} \cite{scipy}, \texttt{Astropy}, \cite{astropy} and \texttt{GetDist} to analysed the MCMC chains developed for COSMOMC \cite{cosmomc}.


\bibliography{./qso_stack}{}
\bibliographystyle{JHEP.bst}
\end{document}